\documentclass[aps, prd, twocolumn, superscriptaddress, nofootinbib, preprintnumbers]{revtex4}
\usepackage[dvipdfmx,hiresbb]{graphicx} 
\usepackage{graphicx}
\usepackage{amsmath}
\usepackage{mediabb}                     
\usepackage{color}   

\allowdisplaybreaks[1]
\usepackage{ulem} 
 
\newcommand{\red}[1]{{\color{black}#1}}
\newcommand{\blue}[1]{{\color{black}#1}}
\newcommand{\magenta}[1]{{\color{black}#1}}

\newcommand{\al}[1]{\begin{align}#1\end{align}}

\usepackage{amsmath,amssymb}
\usepackage{epsf}

\newcommand{\GeV}{\ensuremath{\,\text{GeV} }}
\newcommand{\eV}{\ensuremath{\,\text{eV} }}

\newcommand{\nn}{\nonumber\\}

\newcommand{\infla}{\ensuremath{\text{inf}} }

\begin{document}
\preprint{KUNS-2580}
\title{
\red{
Reheating-era leptogenesis
}
}  
\date{\today}

\author{Yuta Hamada}
\affiliation{
Department of Physics, Kyoto University, Kyoto 606-8502, Japan
}
\author{Kiyoharu Kawana}
\affiliation{
Department of Physics, Kyoto University, Kyoto 606-8502, Japan
}




\begin{abstract}
We propose a novel leptogenesis scenario at the reheating era.
Our setup is minimal in the sense that, in addition to the standard model Lagrangian, we only consider an inflaton and higher dimensional operators. 
The lepton number asymmetry is produced not by the decay of a heavy particle, but by the scattering between the standard model particles.
After the decay of an inflaton, the model is described within the standard model with higher dimensional operators.
The Sakharov's three conditions are satisfied by the following way.
The violation of the lepton number is realized by the dimension-5 operator.
The complex phase comes from the dimension-6 four lepton operator.
The universe is out of equilibrium before the reheating is completed.
It is found that the successful baryogenesis is realized for the wide range of parameters, the inflaton mass and reheating temperature, depending on the cutoff scale. 
Since we only rely on the effective Lagrangian, our scenario can be applicable to all mechanisms to generate neutrino Majorana masses.
\end{abstract}
\maketitle

\normalsize
The discovered Higgs mass $125\GeV$ indicates that the standard model(SM) couplings are perturbative and the electroweak vacuum is (meta)stable even if the cutoff scale of the SM is around the Planck scale~\cite{SM stability,Bare mass}.
The Higgs flat potential around the Planck scale may imply the connection between physics at the electroweak scale and at the Planck scale~\cite{Higgs inflation,criticality}.
Furthermore, the LHC does not find any significant deviations from the SM so far.
These suggest the possibility that the SM is valid up to the very high scale.
However, we know that there exists phenomena which can not be explained within the SM, one of which is the baryon asymmetry of the universe~\cite{Ade:2015xua}
\al{
{n_B\over s}
\simeq
(8.67\pm0.05)\times 10^{-11},
}
where $n_B$ is the number density of the baryon asymmetry and $s$ is the entropy density of the universe.
Taking into account these facts, it is important to consider the simple mechanism to realize the above value under the assumption that the SM is valid up to the very high scale.

There are many proposals to create the baryon asymmetry such as GUT baryogenesis~\cite{Yoshimura:1978ex}, Affleck Dine  baryogenesis~\cite{Affleck:1984fy}, electroweak baryogenesis~\cite{Kuzmin:1985mm}, baryogenesis via leptogenesis~\cite{Fukugita:1986hr}, string scale baryogenesis~\cite{Aoki:1997vb} and so on.

Among the various possibilities,
one of the most simple scenario is leptogenesis in which heavy new particles are introduced, and
the lepton number is produced by the decay of them.
Then, the lepton number is converted into the baryon number by the non-perturbative effect of the SM, namely, the sphaleron process~\cite{'tHooft:1976fv,Klinkhamer:1984di}.
The heavy particles are created by the thermal production~\cite{Fukugita:1986hr} or decay of an inflaton~\cite{Lazarides:1991wu} at the early stage of the universe. 
For example, a heavy particle is the right handed neutrino, triplet scalar~\cite{Hambye:2003ka} or triplet fermion~\cite{Albright:2003xb}.

%
In this letter, we propose a novel leptogenesis scenario. 
In contrast to the conventional scenario, we consider the case where the mass of a heavy particle is larger than the reheating temperature $T_R$.
This situation would be reasonable because it may be natural that the mass of a heavy particle is around the GUT scale, $\mathcal{O}(10^{15})\GeV$, and it is not easy to obtain such a high reheating temperature.
\magenta{
In such a situation, because the on-shell heavy particle is not produced, the lepton asymmetry can not be generated by the decay of the heavy particle.
However, as we will see in the following, by assuming \magenta{an existence of} an inflaton which mainly decays to the SM particles, especially leptons, we show that the lepton asymmetry can be generated by the scattering between the SM particles at the reheating era.
}
%
%

Because the energy scale after the inflation is always smaller than the mass of a heavy particle, we can use the effective Lagrangian:
{\small
\al{\label{effective action}
\mathcal{L}&=
\mathcal{L}_{SM}
+
{1\over \Lambda_1} \lambda_{1,ij} H H \bar{L_j^c} L_i
+
{1\over \Lambda_2^2}\lambda_{2,ijkl}
(\bar{L_i}\gamma^\mu L_j)(\bar{L_k}\gamma_\mu L_l)
\nn
&\phantom{=}+
{1\over \Lambda_3^2}\lambda_{3,ijkl}
(\bar{L_i}\gamma^\mu L_j)(\bar{E_k}\gamma_\mu E_l)
+h.c.
\nn
&\ni
\blue{\lambda_{1,ij}}
{v^2\over 2\Lambda_1}  
(\bar{\nu^c} \nu
+\bar{\nu}\nu^c
)
,}}
where $v\simeq246\GeV$, $H$ is the Higgs doublet, $\mathcal{L}_\text{SM}$ represents the SM Lagrangian, $L$ is the lepton doublet, E is the right handed lepton, and $\nu$ represents the left handed neutrino.\footnote{
By performing the Fierz transformation, the $\Lambda_2$ term is equivalent to 
\al{
{1\over \Lambda_2^2}\lambda_{2,ijkl}
(\bar{L_i}\gamma^\mu L_j)(\bar{L_k}\gamma_\mu L_l)
=2
{1\over \Lambda_2^2}\lambda_{2,ijkl}(\bar{L}_{\red{l}}^cL_{\red{j}}) (\bar{L}_{\red{i}} L^c_{\red{k}}).
}
Therefore, the action here is equivalent to that of Ref.~\cite{Aoki:1997vb} although the situation is different. 
The decay of the higher excited state of string theory is considered in Ref.~\cite{Aoki:1997vb}.
The other study of baryogenesis by the collision can be found in Ref.~\cite{Bento:2001rc}(see also Ref.~\cite{Baldes:2014gca}), where the mirror world is introduced and our scenario is different from their one.
}
Although there are a lot of dimension-6 operators~\cite{Grzadkowski:2010es}, we only consider one which is relevant for our discussion below.
We can make $\lambda_{1,ij}$ real diagonal by the unitary transformation of $L_i$.
In general, $\lambda_{2,ijkl}$ and $\lambda_{3,ijkl}$ become complex in this basis, and therefore break CP .
Although the specific values of $\Lambda_1$, $\Lambda_2$ and $\Lambda_3$ depend on ultraviolet models, $\Lambda_1$ can be written as a function of the neutrino mass if the neutrino is Majorana particle: 
%
%
\al{
\Lambda_1\simeq\magenta{6.0}\times 10^{14}\GeV
\left(
{0.1\eV \over m_\nu}
\right),
}
%
where we have assumed $\lambda_{1,11}\simeq \lambda_{1,22}\simeq\lambda_{1,33}\simeq1$ and $m_\nu:=m_{\nu, 1}\simeq m_{\nu, 2}\simeq m_{\nu, 3}$. 
%
%
%


\blue{Before going into details, let us briefly summarize our scenario here. At the end of the inflation, the inflaton starts to oscillate and to decay into the SM leptons.
The reheating process is completed when the energy density of the SM thermal plasma becomes larger than that of inflaton.
Before the completion of reheating, there are two populations of leptons, one from the inflaton decay, and one in the thermal plasma.
We consider the scattering process between them by the dimension-5 and 6 operators in Eq.(\ref{effective action}).
%
This process is inevitably out of equilibrium, and violates the CP and lepton number.\footnote{
More concretely, in our scenario, it is important that the left handed leptons produced by an inflaton decay are out of equilibrium.
The leptons just after the inflaton decay are different from thermal distribution.
For example, in general, the decay of the inflaton to leptons is flavor non-universal, and the population of leptons are different from each flavor, which plays crucial role to give nonzero asymmetry.
We will see this in Eq.~\eqref{Eq:efficiency2}.
Through the thermalization process of leptons, we obtain the lepton asymmetry of the universe.
}
In particular, due to the CP phase in dimension-6 operator, the interaction \magenta{rate} of $LL\rightarrow \bar{H}\bar{H}$ can differ from \magenta{that of} $\bar{L}\bar{L}\rightarrow HH$. 
Thus, the Sakharov's three conditions are all satisfied, and we can obtain the lepton asymmetry. 
If the reheating temperature is low enough, the washout process can be avoided.
This asymmetry is converted to the baryon asymmetry through the sphaleron process:
\al{
{n_B\over s}
\simeq
{28\over79}{n_L\over s}.
}
}
%

\begin{figure}
\begin{center}
\includegraphics[width=.35\textwidth]{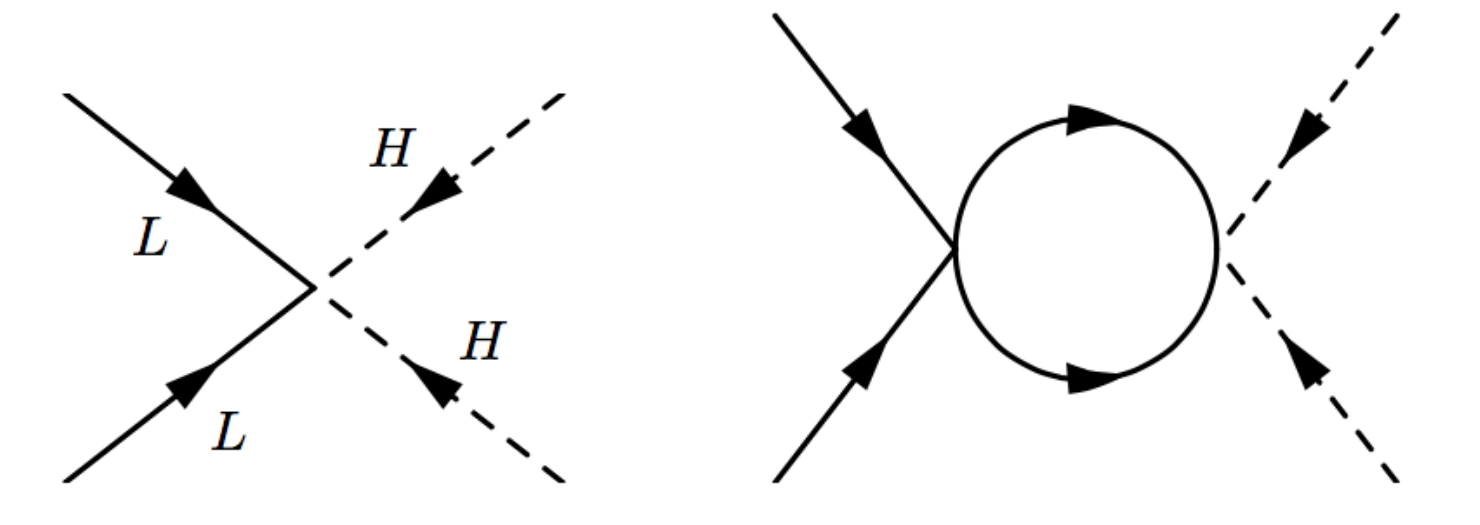}
\end{center}
\caption{The diagram contributing to the lepton asymmetry.
		\red{The solid and dotted lines represent the lepton and Higgs doublet, respectively.}
		}
\label{fig:leptogenesis}
\end{figure}

\begin{figure}
\begin{center}
\includegraphics[width=.4\textwidth]{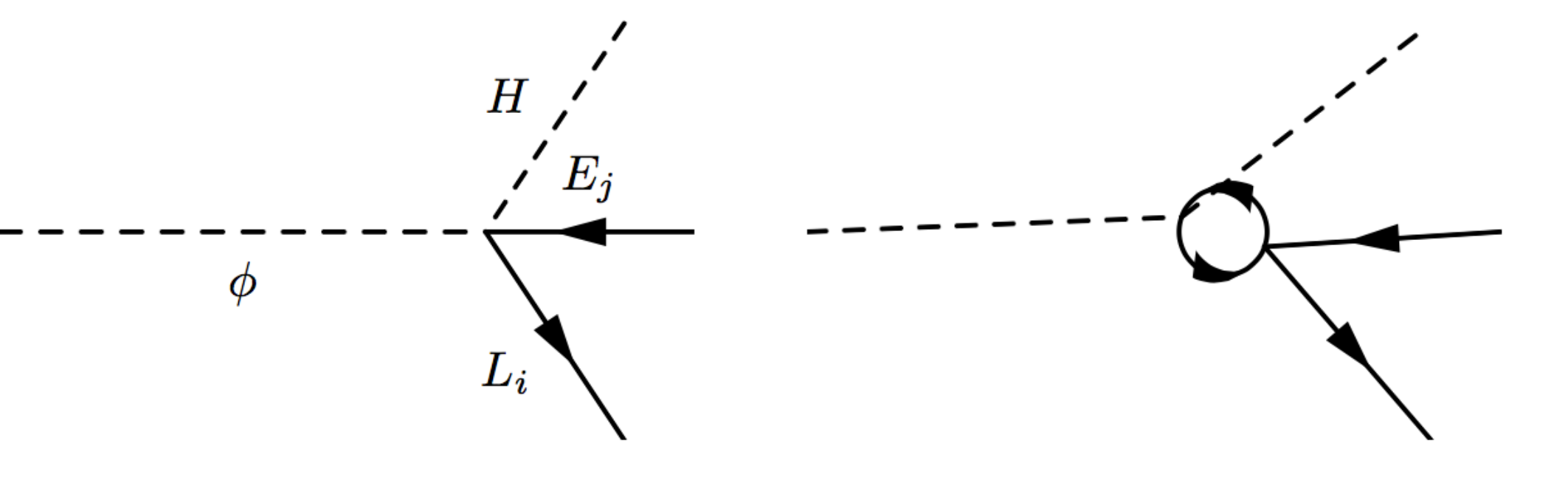}
\end{center}
\caption{The decay of an inflaton.
The interference between the tree and the one-loop diagram induces the CP asymmetry.
\red{
Here the dotted line corresponds to the inflaton.
$i, j$ are the flavor indexes.
}
}
\label{fig:inflaton decay}
\end{figure}

In \blue{this} scenario, there are two kinds of contributions to the lepton asymmetry.
First contribution is that, while the decay of an inflaton itself does not lead to any asymmetry, the scattering between the SM particles induces both of the lepton number and CP violation, see Fig.~\ref{fig:leptogenesis} for the diagram contributing to the lepton asymmetry.
Second contribution is that the CP violation is induced by the decay of an inflaton into \blue{the} lepton pairs (see Fig.~\ref{fig:inflaton decay}), and the lepton number violation is occurred by the succeeding dimension-5 
\blue{operator in Eq.(\ref{effective action})}.



Let us now evaluate the amount of the lepton (baryon) asymmetry.
As for the first contribution, $n_L$ is roughly given by
\al{\label{rough estimation}
{n_L\over s}
\sim
{n_{\text{inf}}\over s} \sum_i 
2\epsilon_i\mathrm{Br}_i
{\Gamma_{\not\!L,i}\over\Gamma_\text{brems}}.
}
\blue{The meaning of \red{each} of the factors in Eq.~\eqref{rough estimation} is as follows.}
The first factor is \blue{the} abundance of an inflaton, which can be written as
\al{
{n_{\text{inf}}\over s} 
&\simeq
{3\over4}{T_R\over m_\text{inf}}
\quad
}
\blue{by equating the energy density of the inflaton $m_\text{inf}n_{\text{inf}}$ with that of the radiation $ 3sT/4$ at the reheating.\footnote{
\magenta{
The same estimation is used, e.g., in the context of leptogenesis by the decay of an inflaton~\cite{Lazarides:1991wu}.
}
}
}
Second, $\epsilon_i$ represents the efficiency
\al{\label{efficiency}
\epsilon_i
:=
2\,
{\sigma_{\bar{L}_i\bar{L}_i\rightarrow HH}
-\sigma_{L_iL_i\rightarrow \bar{H}\bar{H}}\over \sigma_{\bar{L}_i\bar{L}_i\rightarrow HH}+\sigma_{L_iL_i\rightarrow \bar{H}\bar{H}}},
}
where $\sigma$ represents the cross section of each process.
This can be written by
\al{\label{Eq:efficiency2}
\epsilon_i
&\simeq
\sum_j
{1\over2\pi}
{12m_\infla T_R\over \Lambda_2^2}
{\lambda_{1,jj}
\mathrm{Im}(\lambda_{2,ij})
\over
\lambda_{1,ii}
}
,
\quad
}
where $\lambda_{2,ij}:=\lambda_{2,i j i j}$. \blue{Here, note that $\Lambda_1$ dependence is canceled between the numerator and denominator.}

Next, $\mathrm{Br}_i$ denotes the branching ratio of an inflaton to $L_i \bar{L}_i$.
We note that the baryon asymmetry vanishes if we take $\mathrm{Br}_1=\mathrm{Br}_2=\mathrm{Br}_3$
, \footnote{
In this case, Eq.(\ref{rough estimation}) is proportional to
\al{
\sum_{i,j}\lambda_{1,ii}\lambda_{1,jj}\text{Im}\lambda_{2,ij}=\frac{1}{2}\sum_{i,j}\lambda_{1,ii}\lambda_{1,jj}\left(\lambda_{2,ijij}-\lambda^*_{2,ijij}\right).
\label{eq:vanish}
}
However, in Eq.(\ref{effective action}), by including the $h.c$ term and redefining $\lambda_2$, we can show $\lambda_{2,ijij}=\lambda_{2,jiji}^*$. Thus, we can see that the right hand side in Eq.(\ref{eq:vanish}) vanishes.
}
and therefore we simply assume that this is not the case.
Indeed, there is no reason to consider $\mathrm{Br}_1=\mathrm{Br}_2=\mathrm{Br}_3$.
In the following, we take $\mathrm{Br}:=\mathrm{Br}_1\neq0, \mathrm{Br}_2=\mathrm{Br}_3=0$ for simplicity.\footnote{
\red{
We also assume that a vanishing branching ratio of inflaton to $HH$ for simplicity.
It would be interesting to investigate our scenario with general decay mode.
}
}

Finally, $\Gamma_{\not\!L,i}$ represents the interaction rate of the lepton violation process, 
\al{\label{lepton number violation}
\Gamma_{\not\!L,i}
&
\simeq
\magenta{
{11\over4\pi^3}
\zeta(3)
{m_{i, \nu}^2\over v^4}
T_R^3
},
}
\magenta{where $T_R^3$ comes from the number density of the thermal plasma.
On the other hand, } 
$\Gamma_\text{brems}$ is the interaction rate of the thermalization process without the lepton number violation,
\al{\label{rate of brems}
\Gamma_\text{brems}
&\sim
{\alpha_2^2}T_R \sqrt{T_R\over m_\text{inf}}.
}
Here $\alpha_2$ is the $SU(2)_L$ structure constant.
%
%
\red{
If we naively estimate the bremsstrahlung diagram with t-channel gauge boson exchange, we have ${\alpha_2^2}T_R$.
However, because the emission process continues until their energy becomes comparable with thermal bath, the interference among the emission processes should be taken into account.
This interference effect suppresses the thermalization process, which is represented by $\sqrt{T_R/m_\text{inf}}$, and called Landau-Pomeranchuk-Migdal effect~\cite{LPM effect}.
}
Note that 
${\Gamma_{\not\!L}/\Gamma_\text{brems}}$
corresponds to the probability that the lepton number violating process occurs during the time interval $\Delta t=1/\Gamma_\text{brems}$, which is the typical time scale the high energy leptons lose their energy. \footnote{
This statement is valid if the Hubble parameter is sufficiently smaller than $\Gamma_\text{brems}$. In this case, the high energy leptons mainly lose their energy by bremsstrahlung process.
On the other hand, if the Hubble parameter is larger than the $\Gamma_\text{brems}$, the leptons mainly lose the energy by redshift, and the probability is given by ${\Gamma_{\not\!L}/H}$.
The typical values of parameters presented in Eq.(\ref{analytic}) correspond to the former case:
\al{
\Gamma_\text{brems}/H\sim
{\alpha_2^2 M_{pl} \over\sqrt{T_Rm_\text{inf}}}.
}
}
%
\smallskip

As a result, we obtain
{\footnotesize
\al{\label{analytic}
{n_B\over s}
&\simeq
8.7\times10^{-11}
\left({4\times10^{-4}\over\alpha_2^2}\right)
\left({m_\text{inf}\over2\times10^{13}\GeV}\right)^\red{{1\over2}}
\left({T_R\over\red{\magenta{3\times}10^{11}\GeV}}\right)^\red{{7\over2}}\nn
&\times
\left({m_\nu\over0.1\eV}\right)^2
\left({10^{15}\GeV\over\Lambda_2}\right)^2
\left({\mathrm{Br}\times\sum_j \lambda_{1,jj}\mathrm{Im}(\lambda_{2,1j})/\magenta{\lambda_{1,11}}\over2}\right).
}}
We can see that the observed baryon asymmetry is successfully generated.

Here we comment on the possible washout effect.
The interaction rate of the washout process is \magenta{similar to Eq.(\ref{lepton number violation}):}
\al{
\Gamma_\text{wash}
\simeq
{11\over4\pi^3}
\zeta(3)
{\sum m_\nu^2\over v^4}
T^3.
}
From this, one can conclude that the strong wash out is avoided if $T_R$ is sufficiently small.\footnote{
\magenta{
At the time of reheating, the ratio $\Gamma_{\text{wash}}/H$ is 
\al{
\Gamma_{\text{wash}}/H\sim {T_R M_{pl}\over\Lambda_1^2}.
}
If we take $T_R=3\times 10^{11}\GeV$ as a successful example, the ratio is small enough to ignore the washout effect.
%
Note that this does not mean that  the lepton number violation process does not occur.
As clarified in footnote 5, since $\Gamma_\text{brems}\gg H$, the asymmetry is roughly proportional to $\Gamma_{\not\!L}/\Gamma_\text{brems}$, which can be large enough to realize observed baryon asymmetry.
}
}
We emphasize that this washout process is collision between the particles in thermal plasma while Eq.~\eqref{lepton number violation} corresponds to the scattering between leptons in the thermal plasma and ones from inflaton decay. 

\begin{figure}
\begin{center}
\hfill
\includegraphics[width=.23\textwidth]{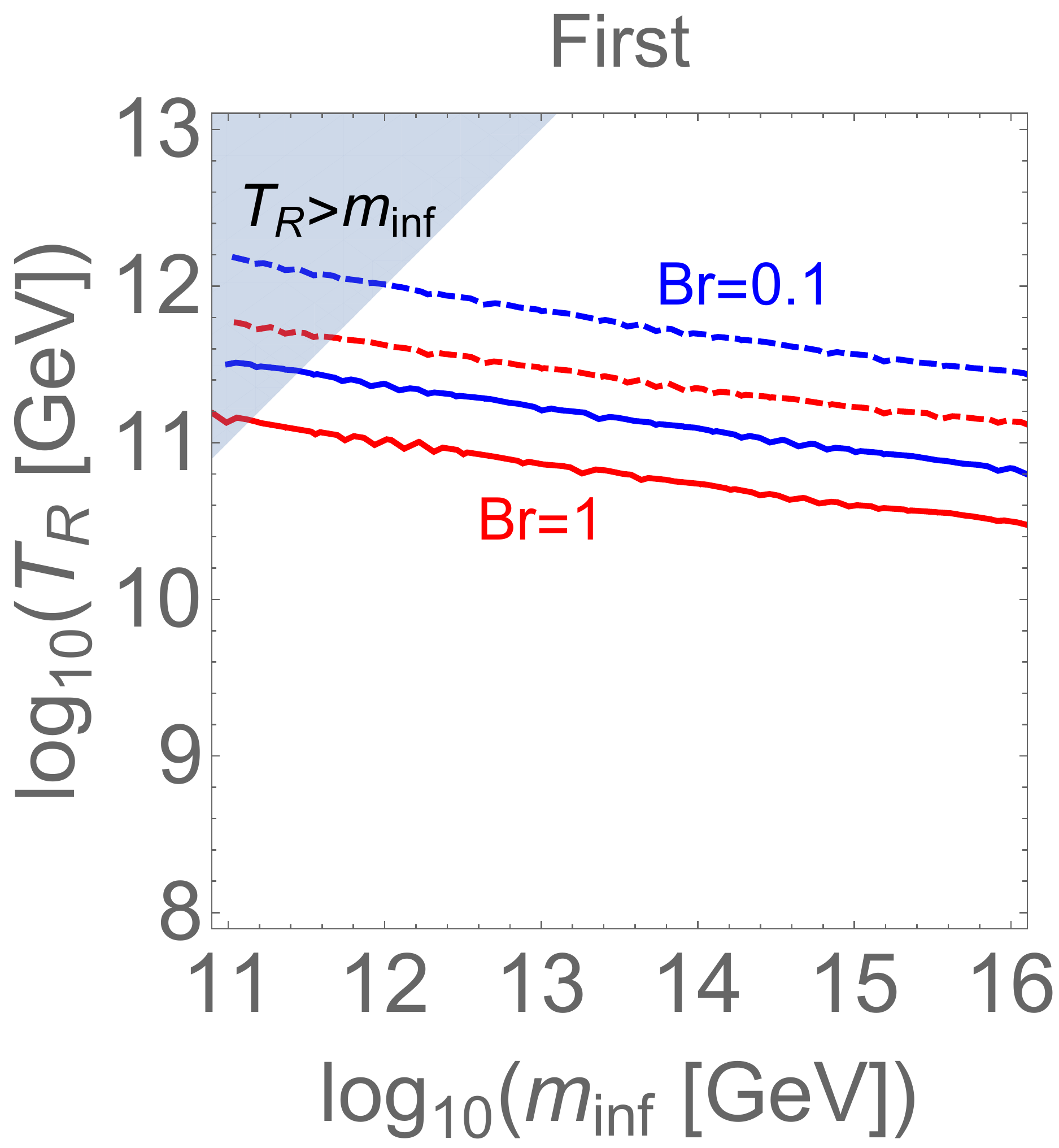}
\hfill
\includegraphics[width=.23\textwidth]{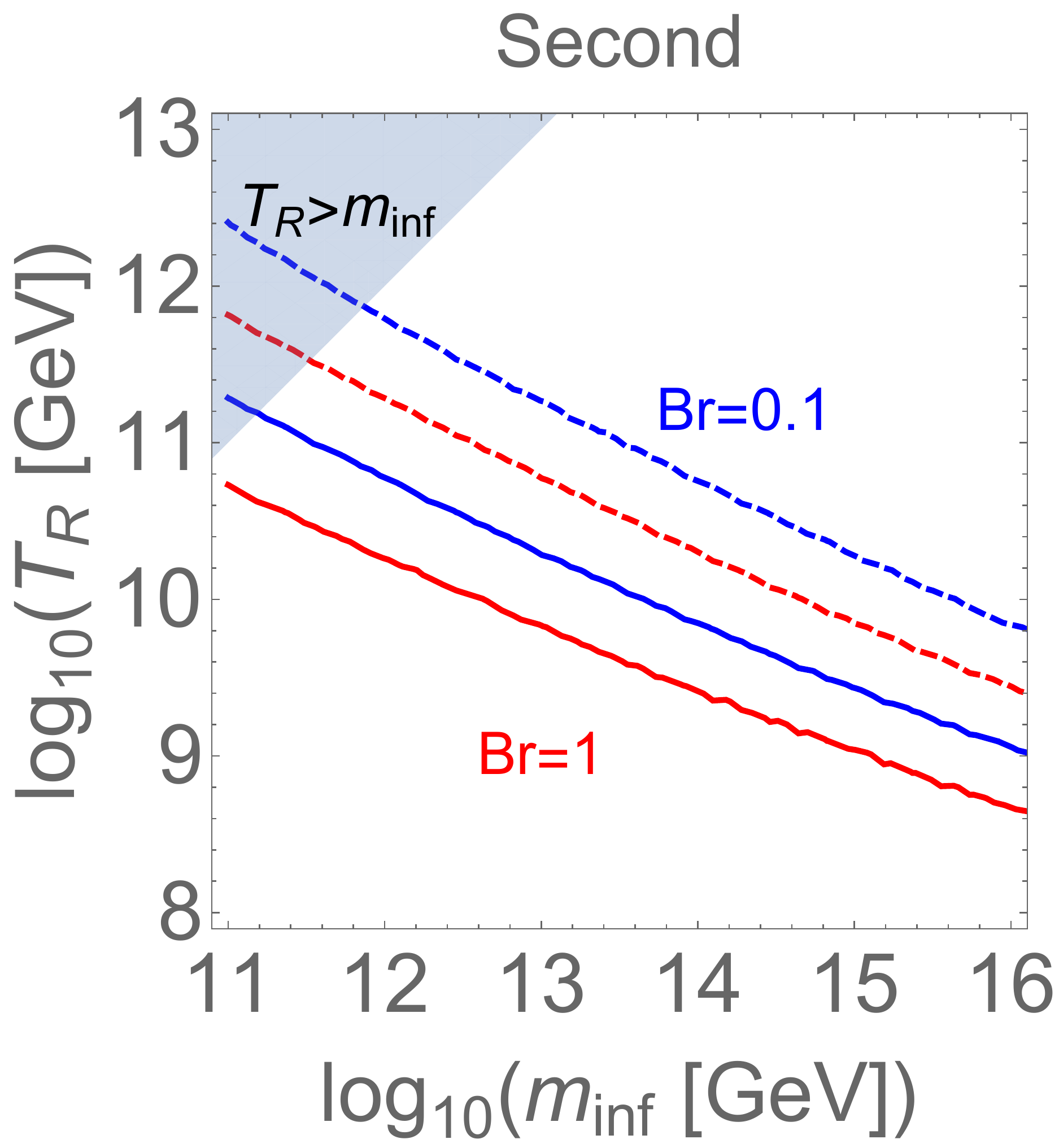}
\hfill\mbox{}
\end{center}
\caption{
Left:
The contour where the first contribution(Eq.\eqref{Boltzmann1}) can explain the observed baryon asymmetry.
Here we take \blue{$\Lambda_\text{inf}=10^{15}$GeV,} $\sum_j \lambda_{1,jj}\mathrm{Im}(\lambda_{2,1j})=2$ and $m_\nu=0.1\eV$.
%
\magenta{
Note that 
the normalization of $a$ can be chosen freely, and we take $a(t=t_\text{end})=1$.
We numerically confirmed that the asymmetry is mainly produced at the time where the reheating is just completed, $t\sim\Gamma_\text{inf}^{-1}$, and therefore the result is insensitive to the value of $\Lambda_\text{inf}$ as long as $T_R\lesssim\Lambda_\text{inf}$.
}
The solid and dashed lines correspond to $\Lambda_2=10^{14}\GeV$ and $\Lambda_2=10^{15}\GeV$, respectively.
The red and blue lines represent $\mathrm{Br}=1$ and $0.1$.
In the blue region, because $T_R>m_\infla$, we need more careful study of the reheating, which is beyond scope of this paper.
Right:
Same as left figure, but considering the second contribution(Eq.\eqref{Boltzmann2}), taking $\sum \mathrm{Im}(y_{1i}^* y_{kl} \lambda_{2,1ikl})/|y_{1i}|^2=1$ and assuming $\lambda_{1,11}^2\gg\lambda_{1,22}^2, \lambda_{1,33}^2$.
}
\label{fig:TvsM}
\end{figure}

In order to confirm above estimation, we solve the following Boltzmann equations:\footnote{We include the effect of redshift as last terms in the right hand side in the second and fourth equations. 
The leptons produced by the inflaton are highly relativistic, and lose their energy by thermalization process or redshift due to the cosmic expansion.
Although the effect of redshift is not important in practice since $\Gamma_\text{brems}>H$ for typical parameters, we include it for completeness.
}
\al{\label{Boltzmann1}
H^2
&={1\over3M_{pl}^2}
\left(
\rho_\infla+{\pi^2 g_*\over30}T^4+{m_\infla\over2} n_l
\right),
\nn
\dot{\rho}_R+4H\rho_R
&=
(1-\mathrm{Br})\Gamma_\infla \,\rho_\infla+{m_\infla\over2}n_l\left(\Gamma_\text{brems}+H\right),
\nn
\dot{n}_L+3 H n_L
&= 
\Gamma_{\not\!L}\,2\epsilon\, n_l
-\Gamma_\text{wash} n_L
,
\nn
\dot{n}_l+3 H n_l
&=
{\Gamma_\infla \,\rho_\infla\over m_\infla}\mathrm{Br}
-n_l
(\Gamma_\text{brems}+H)
,
\nn
\rho_\infla
&=
\Lambda_\infla^4
\left({a(t=t_\text{end})\over a}\right)^3
e^{-\Gamma_\infla t},
}
where 
dot represents the derivative respect to time $t$;
$H$ is the Hubble parameter;
$\rho_\infla$ is the energy density of an inflaton;
$g_*$ is the effective degrees of freedom in the SM;
$T$ is the temperature of radiation;
$\rho_R=\pi^2g_*T^4/30$ is the energy density of radiation;
$n_l$ is the number density of left handed lepton with energy $m_\infla/2$ produced by the decay of an inflaton;
$\Gamma_\infla$ is the decay rate of an inflaton
which is related to the reheating temperature as $T_R\sim \sqrt{M_{pl}\Gamma_{\text{inf}}}$
;
$\Lambda_\infla^4$ is the energy density at the end of inflation;
$a$ is the scale factor;
$t_\text{end}$ is the time when inflation ends.
%
%
We have checked that the numerical results agree with Eq.~\eqref{analytic} within one order of magnitude.
See the left panel of Fig.~\ref{fig:TvsM}.

\begin{figure}
\begin{center}
\includegraphics[width=.17\textwidth]{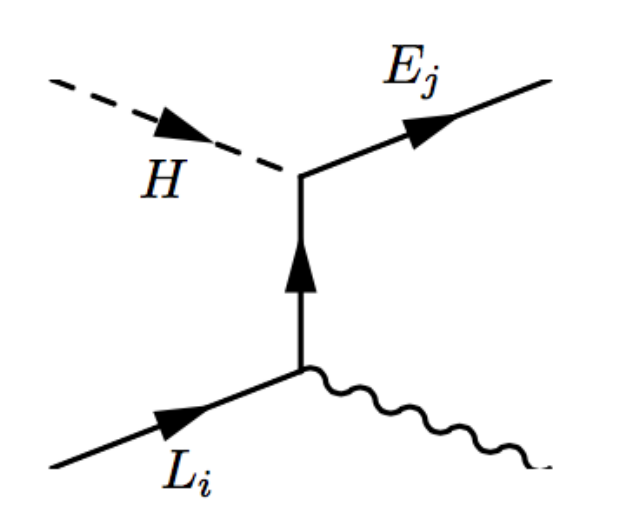}
\end{center}
\caption{
\red{
The washout process of lepton flavor asymmetry.
$E$ is the right handed charged lepton field.
}
}
\label{fig:washout}
\end{figure}

Now let us move to the estimation of the second contribution to the lepton asymmetry.
First, the asymmetry between left and right lepton is occurred by the decay of an inflaton due to the interference between tree and one-loop decay process, see Fig.~\ref{fig:inflaton decay}.
Succeedingly, the lepton number is violated by the dimension-5 operator.\footnote{The second contribution is different from the leptogenesis by inflaton decay~\cite{Lazarides:1991wu} with the strong washout, in which the heavy Majorana neutrinos, produced by the inflaton decay, present both of lepton number and CP violations. 
The lepton number violating SM scattering just decreases the total number of asymmetry.
On the contrary, in our scenario, the decay of heavy particle(inflaton) only provides the CP violation, and the actual lepton asymmetry is produced by the scattering between these SM particles.
}
Since the only left handed leptons feel lepton number violation, we have nonzero lepton asymmetry.
In order to illustrate the idea, we assume that an inflaton mainly decays to lepton pairs by the dimension-5 operator \footnote{
In the case of  three body decay, the energy of the decay product is not exactly $m_\text{inf}/2$, but becomes continuous spectrum.
This would not change the order of asymmetry, but change the factor of asymmetry.
}
\al{
{1\over M}
y_{ij} \phi \bar{L}_i H  E_j
}
where $M$ is some scale, $y_{ij}$ is the coupling, and $\phi$ is the inflaton.\footnote{
Another choice is considering the models where the inflaton has same charge as the SM Higgs field.
Then, the inflaton can decay into right and left lepton pairs by dimension-4 operator, which can become main channel by taking appropriate value of the coupling.
Although such a coupling often threatens the flatness of inflaton potential, it is known that the inflation is possible by introducing non-minimal coupling between gravity and the inflaton sector~\cite{Kallosh:2013tua}\cite{Higgs inflation}.
The argument below is not drastically modified even in this case.
}
By assuming that $\lambda_{1,11}$ is larger than $\lambda_{1,22}$ and $\lambda_{1,33}$, the net lepton asymmetry is given by 
\al{\label{contribution II}
{n_L\over s}
\sim
{n_\infla\over s}\,
\sum_{i=1,2,3}
2
\epsilon_{1,i}\,
\mathrm{Br}_{1i}
{\Gamma_{\not\!L,1}\over\Gamma_\text{LF}},
}
where $\mathrm{Br}_{ij}$ is the branching ratio of an inflaton to $HL_i\bar{E}_j$, $\epsilon_{i,j}$ is
{\footnotesize
\al{
\epsilon_{i,j}
&:=
{\Gamma(\phi\rightarrow \bar{H} L_i \bar{E}_j)-\Gamma(\phi\rightarrow H\bar{L}_i E_j)
\over
\Gamma_\infla}\nn
\label{epsilon2}
&\simeq 
{1\over8\pi}
{m_\infla^2\over\Lambda_3^2}
\sum_{k,l}
{
\mathrm{Im}(y_{ij}^*y_{kl}\lambda_{3,iklj})
\over
|y_{ij}|^2
}
,
}}
and
$\Gamma_\text{LF}$ is the rate of washout of the asymmetry between left and right leptons,
\al{
\Gamma_\text{LF}
\sim
\alpha_2 \alpha_{y_l} T_R
,
}
see also Fig.~\ref{fig:washout}
.
Here \red{$\alpha_{y_l}$ is the structure constant of the charged lepton Yukawa}. 
%
%
Then, we obtain
\red{
{\small
\al{
{n_B\over s}
&
\simeq
8.7\times10^{-11}
\left({2\times10^{-2}\over\alpha_2}\right)
\left({8\times10^{-6}\over\alpha_{y_l}}\right)
\left({m_\text{inf}\over2\times10^{13}\GeV}\right)\nn
&\times
\left({T_R\over4\times10^{10}\GeV}\right)^{3}
\left({m_\nu\over0.1\eV}\right)^2
\left({10^{15}\GeV\over\Lambda_3}\right)^2
\nn
&\times
\left({\sum_{i=1,2,3}\mathrm{Br}_{1i}\over1}\right)
\left(
{\sum \mathrm{Im}(y_{1i}^* y_{kl} \lambda_{3,1kli})/|y_{1i}|^2\over 1}
\right)
.
}
}
}

In this case, Boltzmann equations are the same as Eq.~\eqref{Boltzmann1} except for the third line.
Instead of the third line, we have
{\small
\al{\label{Boltzmann2}
\dot{n}_L+3 H n_L
&= 
-\sum_{j}2\Gamma_{\not\!L,j}n_{Lj}
-\Gamma_\text{wash} n_L
,
\nn
\dot{n}_{L1}-\dot{n}_{Lj}
+3 H (n_{L1}-n_{Lj})
&=
{\Gamma_\infla \,\rho_\infla\over m_\infla}\mathrm{Br}_{1j}
\,2\epsilon_{1,j}
\nn
&\,\,\,\,\,- (n_{L1}-n_{Lj})
\red{\Gamma_\text{LF}}
,
\quad
(j=2,3)
}
}
where $n_{Li}$ is the lepton number asymmetry of $L_{i}$. 
In Fig.~\ref{fig:TvsM}, we show contours which realizes the observed baryon asymmetry.
In this figure, we simply assume that the effective field theory picture is valid at the time of reheating.
This is valid if the mass of inflaton and the reheating temperature are lower than that of new particles which appear in ultraviolet completion.
In order to check whether this is true or not, the model dependent analysis is needed, which we leave for future publication~\cite{Hamada:2016npz}.

%
%
%

Finally we discuss the possible origin of the higher dimensional operators.
For example, in the case of type-I seesaw, the dimension-5 and dimension-6 operators are generated by integrating out the right handed neutrino: 
\al{
&
{\lambda_{1,ij}\over\Lambda_1}\simeq(y_{\nu} m_N^{-1} y_\nu^T)^i_j,
&
{\lambda_{2,ijkl}\over\Lambda_2^2}\simeq(y_{\nu} y_\nu^\dagger)^i_j(y_{\nu} y_\nu^\dagger)^k_l {1\over16\pi^2 m_N^2},
}
where $y_\nu$ is the neutrino Yukawa coupling and $m_N$ is the right handed neutrino mass.
We can see that the value of $\Lambda_2$ is $\mathcal{O}(10^{15})\GeV$ due to the one-loop suppression.
Of course we have different $\Lambda_2$ for other seesaw models. 
In Zee model~\cite{Zee:1980ai},  $\Lambda_2=\mathcal{O}(10^{13})\GeV$ because $\Lambda_1$ term is one-loop suppressed while $\Lambda_2$ term is generated at the tree level. 
In type-II seesaw~\cite{typeII seesaw} and Ma model~\cite{Ma:2006km}, $\Lambda_2=\mathcal{O}(10^{14})\GeV$ since $\Lambda_1$ and $\Lambda_2$ terms are the same order of magnitude.
It would be interesting to examine the value of $\Lambda_2$ in various neutrino models.
As for the dimension-6 operator suppressed by $\Lambda_3$, it can be generated at the tree level, e.g., if the second Higgs doublet heavier than the electroweak scale is added in addition to the SM Higgs doublet.


Before concluding this letter, a few more remarks are given in order. 
Firstly, we emphasize that, since our analysis is only based on the effective Lagrangian, our scenario can be applicable to all mechanisms to generate neutrino Majorana masses.
Secondly, although we do not consider in this paper, if the right handed neutrino is lighter than inflaton, it may be produced by the scattering between the SM particles.
It may be interesting to investigate whether this scenario can explain the baryon asymmetry or not.
Thirdly, 
our scenario goes well with the high scale inflation model.
In the case of the quadratic chaotic inflation~\cite{Linde:1983gd}, we have $m_\infla\simeq2\times10^{13}\GeV$. 
The observed baryon asymmetry is explained if the inflaton couples with the SM particles by the dimension-5 operator, $\Gamma_\infla\simeq{1\over8\pi}{m_\infla^3\over M^2}{1\over100}$,
where $M\sim 10^{15\text{--}16}\GeV$.
It is remarkable that $M$ is close to the string/GUT scale, and that such a simple high scale inflation model actually realizes successful baryogenesis.
It is also interesting to explore if other inflation scenarios such as Higgs inflation are consistent with this mechanism.
\magenta{Furthermore, our scenario might be applicable not only to the decay of an inflaton, but also to that of more general late decaying scalar field such as moduli field.}
Another interesting direction is to examine the flavor effects on our scenario.
Finally, our scenario is excluded if it turns out that the neutrino is Dirac fermion or Majorana mass is very small compared to $\mathcal{O}(0.1)\eV$,
\red{
see also Ref.~\cite{Deppisch:2015yqa}.
}

In conclusion, we have proposed a novel scenario for baryogenesis.
We have shown that, even if the reheating temperature and an inflaton mass are smaller than a heavy particle mass, the observed baryon asymmetry is successfully generated.
%
\red{
Our argument is quite general, and can be applicable to many models. 
We believe that this work innovates the new class of baryogenesis scenario. 
}

\subsection*{Acknowledgement}
We thank \red{Keisuke Harigaya}, Hikaru Kawai, \red{Kyohei Mukaida}, Koji Tsumura and Masaki Yamada for useful discussions and comments.
This work is supported by the Grant-in-Aid for Japan Society for the Promotion of Science (JSPS) Fellows No.251107 (YH) and No.271771 (KK).

\end{document}